\documentstyle[12pt]{article}

\begin{document}

\begin{center}

{\Large{\bf Self--Similar Crossover in Statistical Physics} \\ [5mm]

V. I. Yukalov\footnote{corresponding author} and S. Gluzman$^2$} \\ [3mm]

{\it $^1$ Bogolubov Laboratory of Theoretical Physics \\
Joint Institute for Nuclear Research, Dubna 141980, Russia \\ [2mm]

$^2$ International Centre of Condensed Matter Physics \\
University of Brasilia, Brasilia, DF 70919--970, Brazil}

\end{center}

\vspace{2cm}

\begin{sloppypar}

\begin{abstract}

An analytical method is advanced for constructing interpolation formulae 
for complicated problems of statistical mechanics, in which 
just a few terms of asymptotic expansions are available. The method is 
based on the self--similar approximation theory, being its variant where 
control functions are defined from asymptotic crossover conditions. Several
examples from statistical physics demonstrate that the suggested method 
results in rather simple and surprisingly accurate formulae.

\end{abstract}

\vspace{1cm}


\vskip 2cm


\newpage 

\section{Introduction}

In many problems of statistical physics one encounters the so--called 
crossover phenomena, when a physical quantity qualitatively changes its 
behaviour in different domains of its variable. To be more precise, we 
may specify a crossover as follows. Let a function $f(x)$ represent a 
physical quantity of interest, with a variable running through the 
interval $x_1\leq x\leq x_2$. And let the behaviour of this function, 
describing some physical process, be essentially different near the 
boundary points $x_1$ and $x_2$. Assume that the function varies 
continuously from $f(x_1)$ to $f(x_2)$, as $x$ changes from $x_1$ to 
$x_2$. Then we may say that the function in the interval $[x_1,x_2]$ 
undergoes a crossover between $f(x_1)$ and $f(x_2)$.

Crossover behaviour of different physical quantities is so ubiquitous in
nature that one could list a plenty of examples. For instance, a number 
of physical quantities essentially change their behaviour when passing 
from the weak--coupling to strong--coupling limit [1]. In the theoretical
description of crossover there exists a problem which is common for
practically all physical applications. Real physical systems are usually so
complicated that describing them equations almost never can be solved
exactly. However, it is often possible to find asymptotic expansions of
solutions in vicinity of boundary points. The natural arising problem is how
to construct an accurate approximation for the sought function, valid on the
whole domain of its variable, knowing only its asymptotic behavior near the
boundaries. This problem is aggravated by the fact that only a few terms of
the asymptotic expansions are usually available. In such a case the problem
looks unsolvable.

The most known method of treating the interpolation problem is by using 
the so--called two--point Pad\'e approximants [2-4] or, equivalent to 
the latter, the Thron continued fractions [5,6]. In many cases  the 
two--point Pad\'e approximation yields quite reasonable interpolation 
formulas. However, the usage of this method has not become widespread 
because of the following shortcomings of the Pad\'e approximants:

\vspace{2mm}

(i) When constructing these approximants, one often obtains spurious poles 
yielding unphysical singularities [2-4]. A sequence of Pad\'e approximants 
may even have infinitely many poles [7].

\vspace{2mm}

(ii) A number of examples are known when Pad\'e approximants are not able 
to sum perturbation series even for small values of an expansion parameter 
[8,9].

\vspace{2mm}

(iii) In the majority of cases, except some trivial examples, to reach a 
reasonable accuracy, one needs to have tens of terms in perturbative 
expansions [4]. While, as is emphasized above, in physically interesting 
problems one often has only a few terms.

\vspace{2mm}

(iv) Defining the two--point Pad\'e approximants one always confronts the 
ambiguity in distributing the coefficients, that is, in deciding which of 
the coefficients must reproduce the left--side expansion and which the 
right--side series [1-6]. Such an ambiguity increases with the increase 
of approximants orders, making it difficult to compose two--point Pad\'e 
tables. And for the case of a few terms, this ambiguity makes the 
two--point Pad\'e approximants practically unapplicable. A nice analysis 
of the latter difficulty was done in Ref. [10], where it was shown that, 
for the same problem, one may construct different two--point Pad\'e 
approximants all having correct left--side and right--side limits, but 
differing from each other in the intermediate region by a factor of $40$
which gives $1000\%$ of uncertainty. This clearly demonstrates that in 
the case of short series the two--point Pad\'e approximation not only 
cannot provide a reasonable quantitative approach but even does not permit 
to get a qualitative description. The latter concerns the general situation, 
although there can happen some trivial cases when two--point approximants 
make sense even being built with a few perturbative terms. However, their 
application to such few--term cases, in general, is absolutely unreliable.

\vspace{2mm}

(v) The two--point Pad\'e approximants can be used for interpolating between 
two different expansions not always, but only when these two expansions have 
compatible variables [2-4,9]. When these expansions have incompatible 
variables, the two--point Pad\'e approximants cannot be defined in principle.

\vspace{2mm}

(vi) When interpolating between two points, one of which is finite and 
another is at infinity, one is able to describe at infinity only rational 
powers [2-4]. The impossibility to deal with nonrational powers limits 
the applicability of the two--point Pad\'e approximation.

\vspace{2mm}

(vii) The problem of approximating the functions increasing at infinity 
is especially difficult. A two--point Pad\'e approximant can treat only a 
power--law increase [1-5] and is not able to describe other types of 
behaviour. But in physical problems the functions of interest often 
exhibit at infinity quite different behaviour, for example, growing 
exponentially or following other more complicated ways. In such cases the 
two--point Pad\'e approximants are useless.

\vspace{2mm}

The difficulties listed above are well known and discussed in literature. 
We have cited here only some important references [1-10]. More details on 
mathematical problems in Pad\'e approximation and its applications can be 
found in several volumes of papers, e.g. in Ref. [11].

As follows from the above discussion, the two--point Pad\'e approximation 
in many cases is not applicable. It is evident that there is a necessity 
of developing a more general approach which could overcome the discussed 
difficulties and would be applicable to a larger variety of problems, 
including those for which the two--point Pad\'e approximants cannot be 
used. It is important that such an approach would provide relatively 
simple analytical formulas for the physical quantities of interest. The 
advantage of having analytical expressions, as compared to just numbers 
that could be obtained from a numerical procedure, is in the convenience 
of analysing such expressions with respect to various parameters entering 
into them. Therefore, we keep in mind an analytical, rather than 
numerical, method that would combine relatively simple representations 
for physical quantities with their good accuracy.

It is worth emphasizing that to derive a new physical formula, valid in 
the whole range of physical variables, is not merely a mathematical 
challenge but this provides new physics, since in the majority of cases 
realistic physical problems correspond neither to weak coupling regime 
nor to strong coupling limit, but to the intermediate range of parameters.
Therefore, it is of great importance for physics to possess a general 
mathematical tool permitting to derive explicit crossover formulas for 
arbitrary physical phenomena.  

In the present paper we suggest an approach for treating this problem. Our
approach is based on the self--similar approximation theory [12-22]
permitting an accurate reconstruction of functions from a few terms of
perturbative expansions. The effectiveness of the self--similar approximation
theory is due to the usage of powerful techniques of dynamical theory
and optimal control theory. Fast convergence of renormalized sequences is
achieved by means of control functions. In the algebraic self-similar
renormalization [20-22], we required the algebraic invariance of
renormalization-group flow. Then, control functions are introduced as powers
of a multiplicative algebraic transformation. These control functions are
defined by the stability and fixed-point conditions for a dynamical system
called the approximation cascade. In general, the evolution equations for a
dynamical system can be completed by additional constraints whose existence
imposes restrictions on the definition of control functions.

Crossover problem presents an example when additional constraints appear
absolutely naturally. Really, assume that we have a $k$-order expansion 
$p_k(x)$ approximating the sought function $f(x)$ in the asymptotic vicinity
of the left boundary $x=x_1$. And suppose that we are given an asymptotic
behavior of this function near the right boundary $x=x_2$. For a moment,
take for simplicity that we are given the value $f(x_2)$ at the right
boundary point $x=x_2$. When constructing a self-similar approximation 
$f_k^*(x)$ by renormalizing the left boundary expansion $p_k(x)$, we have
as an additional constraint the right boundary condition 
$f_k^{*}(x_2)=f(x_2)$ .

We show below that the algebraic self-similar renormalization provides a
very convenient tool for treating the crossover problem. This approach
permits us to find, having just a minimal information about the asymptotic
behavior of a function near boundary points, a quite accurate approximation
for the whole region of the variable. In the majority of cases the maximal
error of a self--similar approximation is a few percent and in many cases 
not more than one percent. In addition to being quite accurate, this 
approximation is usually given by very simple expressions that are easy 
to analyze. We illustrate the approach by several examples from different 
branches of statistical physics. The variety of considered cases 
emphasizes the generality of the approach and proves that it is a very 
effective tool for treating arbitrary crossover phenomena.

Recently, we have applied such an interpolation approach to several 
quantum--mechanical problems [23]. However, what makes the latter 
principally different from the problems of statistical physics is that in 
quantum mechanics one usually possesses quite a number of terms of 
perturbative expansions, while in statistical physics this luxury is 
rather rare, so that in the majority of cases one is able to derive just 
a few perturbative terms. In the present paper we aim at showing that our 
interpolation method does work for those complicated problems of 
statistical physics where only a few terms of asymptotic expansions are 
available and other methods are not applicable. Nevertheless, the 
self--similar interpolation makes it possible to treat even such 
complicated crossover problems, obtaining simple and accurate formulae.

\section{General approach}

In this section, we formulate the general scheme of the approach not
specifying the physical nature of a considered function. Let us be interested 
in a function $f(x)$ of a real variable $x$. Assume that in the vicinity 
of some point $x=x_0$ there exist asymptotic expansions $p_k(x,x_0)$, with 
$k=0,1,2,\ldots$, corresponding to this function, 
\begin{equation}
\label{1}
f(x)\simeq p_k\ (x,x_0), \qquad x\rightarrow x_0 \; . 
\end{equation}
Following the algebraic self--similar renormalization procedure [20-22], we
define the algebraic transform 
\begin{equation}
\label{2}
P_k(x,s,x_0)=x^s p_k(x,x_0)\; , 
\end{equation}
where $s$ is yet unknown, and later will play the role of a control
function. The transform inverse to that in Eq.(2) is 
\begin{equation}
\label{3}
p_k(x,x_0)=x^{-s} P_k (x,s,x_0)\; . 
\end{equation}
Then we have to construct an approximation cascade with a trajectory
bijective to the approximation sequence $\left\{ P_k\right\} $. This
procedure with all necessary mathematical foundations and details has been
described in Refs. [13-19]. So, we sketch here only the main steps needed for
grasping the idea of the method and we concentrate on those formulas that
permit us to apply the method for crossover phenomena.

Define an expansion function $x=x(\varphi,s,x_0)$ by the equation 
\begin{equation}
\label{4}
P_0(x,s,x_0) = \varphi ,\qquad  x = x(\varphi,s,x_0) \; , 
\end{equation}
where $P_0$ is the first available term from the sequence $\{ P_k\}$. 
Introduce a function 
\begin{equation}
\label{5}
y_k(\varphi,s,x_0) = P_k(x(\varphi,s,x_0),s,x_0)\; . 
\end{equation}
The transformation inverse to Eq. (5) reads 
\begin{equation}
\label{6}
P_k(x,s,x_0)=y_k(P_0(x,s,x_0),s,x_0)\; . 
\end{equation}
The family of endomorphisms, $\{y_k\}$, forms a cascade with the velocity
field 
\begin{equation}
\label{7}
v_k(\varphi,s,x_0) = y_{k+1}(\varphi,s,x_0) - y_k(\varphi,s,x_0)\; . 
\end{equation}
The trajectory of the cascade $\{y_k\}$ is, by definitions (5) and (6),
bijective to the approximation sequence $\{ P_k\}$. Embedding the 
approximation cascade into an approximation flow [16-19] and integrating
the corresponding evolution equation, we come to the evolution integral 
\begin{equation}
\label{8}
\int_{P_k}^{P_{k+1}^*}\frac{d\varphi}{v_k(\varphi,s,x_0)} = \tau \; , 
\end{equation}
in which $P_k=P_k(x,s,x_0)$ is any given term from the approximation
sequence $\{ P_k\} ;\; P_{k+1}^*=P_{k+1}^*(x,s,\tau,x_0)$ is a 
self--similar approximation representing a fixed point of the
approximation cascade; and $\tau$ is an effective minimal time necessary
for reaching the fixed point.

Recall that we started with a sequence $\{ p_k\}$ of asymptotic
expansions for the considered function $f(x)$. Then we passed to the
sequence $\{ P_k\}$ by means of the algebraic transformation (2). Now we
have to return back employing the inverse transformation (3). To this end,
we set 
\begin{equation}
\label{9}
F_k^*(x,s,\tau,x_0) = x^{-s} P_k^*(x,s,\tau,x_0)\; . 
\end{equation}

The quantities $s$ and $\tau$ are the control functions guarantying the
stability of the method, that is, the convergence of the procedure. These 
functions are to be defined by the stability conditions, such as the 
minimum of multiplier moduli, together with additional constraints, like, 
e.g., boundary conditions. Let us find from such conditions $s=s_k$ and 
$\tau =\tau_k$. Substituting these into Eq.(9), we obtain the self--similar 
approximation 
\begin{equation}
\label{10}
f_k^*(x,x_0) = F_k^*(x,s_k,\tau_k,x_0)\; 
\end{equation}
for the function $f(x)$. We retain here the notation for the point $x_0$ in
order to show that the approximation (10) has been obtained by renormalizing 
$p_k(x,x_0)$ which, according to Eq. (1), is an asymptotic expansion of 
$f(x)$ in the vicinity of the point $x=x_0$.

Now assume that the variable $x$ changes in the interval $x_1\leq x\leq x_2$
and that the asymptotic behavior of a function $f(x)$ is known near the
boundaries of this interval. The latter means that in Eq. (1) we have to put,
instead of $x_0,$ either $x_1$ or $x_2$. Let us take, for concreteness, the
boundary points $x_1=0$ and $x_2\rightarrow\infty$. Then we have two types
of expansions, $p_k(x,0)$ and $p_k(x,\infty)$. Following the procedure 
described above, we can construct, in the place of Eq. (9), two quantities, 
$F_k^*(x,s,\tau,0)$ and $F_k^*(x,s,\tau,\infty)$.

As is discussed above, the control functions $s$ and $\tau$ are to be
defined from stability conditions plus additional constraints. The natural
such constraints for the crossover problem can be formulated as follows.
Suppose we have constructed the renormalized expression $F_k^*(x,s,\tau,0)$ 
starting from the left asymptotic expansion $p_k(x,0)$. By this
construction, the function $F_k^{*}$ has correct asymptotic behavior near
the left boundary. But in order to correctly represent the sought function
in the whole interval of $x\in [0,\infty)$, the renormalized expression 
$F_k^{*}$ must have the correct asymptotic behavior when approaching the
right limit. This implies the validity of the condition 
\begin{equation}
\label{11}
\lim_{x\rightarrow\infty} \left| F_k^*(x,s,\tau,0) - p_i(x,\infty)
\right| = 0\; , 
\end{equation}
imposing constraints on $s=s_k$ and $\tau=\tau_k$. We shall call Eq. (11) 
providing the correct crossover behavior from the left to the right 
boundary the {\it left crossover condition.} The quantity $s_k$
can be called the {\it left crossover index,} and $\tau_k$, the {\it 
left crossover time}. For the self--similar approximation (10) we get, in
this way, 
\begin{equation}
\label{12}
f_k^*(x,0) = F_k^*(x,s_k,\tau_k,0)\; , 
\end{equation}
which may be named the {\it left self--similar approximation}, or the
{\it left crossover approximation.}

The analogous reasoning works, as is clear, when we are considering the
crossover from the right to left. Then we obtain the {\it right crossover
condition}
\begin{equation}
\label{13}
\lim_{x\rightarrow 0} \left| F_k^*(x,s,\tau,\infty) - p_j(x,0)
\right| = 0\; , 
\end{equation}
imposing constraints on $s=s_k$ and $\tau =\tau_k$, thus defining
the {\it  right crossover index} $s_k$ and the {\it right crossover time} 
$\tau_k$. As a result, we come to the {\it right self--similar approximation},
or the {\it right crossover approximation}
\begin{equation}
\label{14}
f_k^*(x,\infty) = F_k^*(x,s_k,\tau_k,\infty)\; . 
\end{equation}
In general, from the left and the right approximations, (12) and,
respectively, (14), we can compose the {\it average self--similar
approximation}, or {\it average crossover approximation} 
\begin{equation}
\label{15}
f_k^*(x) = \frac{1}{2} \left [ f_k^*(x,0) + f_k^*(x,\infty)
\right ] \; . 
\end{equation}

The suggested general approach to reconstructing crossover functions can be
employed for any crossover phenomena. In particular applications, it can
happen that we possess a reliable asymptotic expansion only from one side of
the crossover domain, and from another side just one term is available. In
this case, as is clear, we are not able to construct both left and right
self-similar approximations, but only one of them. Nevertheless, such
one--side approximations are usually quite accurate, as we show by 
examples in the following sections. The possibility of constructing
accurate approximations, when we have a perturbative series only from one
side of the crossover region and a sole asymptotic term from another side,
is very important since this situation constantly occurs in realistic
physical problems. We shall demonstrate in what follows how it is possible
to improve the accuracy of such one--side approximations by combining the
terms of a given one--side series and defining the crossover indices so that
to satisfy the asymptotic limit from another side, in accordance with the
crossover conditions (11) or (13).

In order to emphasize that the suggested approach does work even for the
cases with a very scarce information about the sought function, let us
consider a simple example. Suppose that we know the asymptotic behavior of a
function near the left boundary, where $x\rightarrow 0$, only in the linear
approximation
\begin{equation}
\label{16}
p_1(x,0)\simeq a_0 + a_1 x\; , \qquad  a_0,\; a_1 \neq 0\; . 
\end{equation}
And assume that only one asymptotic term is known from the right side, 
\begin{equation}
\label{17}
p_1(x,\infty) \simeq A~x^n\; , \qquad A,\; n \neq 0\; , 
\end{equation}
as $x\rightarrow\infty$. In such an extreme case of minimal information,
it looks like there is no a regular way of recovering the function for the 
whole axis $0\leq x< \infty $. However, our approach, based on the idea of
self--similarity, permits us to recover the sought function.

Following the procedure described above, in the place of Eq.(9), starting
from expansion (16), we obtain
\begin{equation}
\label{18}
F_1^*(x,s,\tau,0) = a_0 \left( 1- \frac{a_1\tau }{a_0s}~x
\right)^{-s} \; . 
\end{equation}
With Eqs.(17) and (18), the left crossover condition (11) reads
$$
\lim_{x\rightarrow \infty} \left | a_0 \left ( 1- 
\frac{a_1\tau}{a_0s}~x \right)^{-s} - A~x^n \right | = 0\; , 
$$
From here the left crossover index $s_1$ and the left crossover time 
$\tau_1$ are uniquely defined as follows: 
\begin{equation}
\label{19}
s_1 = -n \; , \qquad \tau_1 = n \frac{a_0}{a_1} \left ( 
\frac{A}{a_0} \right )^{1/n} \; . 
\end{equation}
Substituting these values into Eq. (18), as is prescribed by Eq. (12), we
recover the left crossover approximation 
\begin{equation}
\label{20} 
f_1^*(x,0) = a_0 \left [ 1 + \left ( \frac{A}{a_0} \right )^{1/n}
x \right ]^n \; . 
\end{equation}
At large $x\rightarrow\infty$, expression (20) reduces to the limit (17).
When $x\rightarrow 0$, then we have the linear behavior
$$
f_1^*(x,0) \simeq a_0 + a_1^* x\; , 
$$
where $a_1^*=na_0\left( A/a_0\right)^{1/n}$ is the renormalized
coefficient. Such a renormalization is typical of renormalization group
techniques, as is discussed in Refs. [20-22].

Thus, even having so scanty information about the asymptotic properties of a
function, as in the above example, our approach allows us to reconstruct, in 
a systematic way, the function for the whole domain of its variable. This 
reconstruction becomes possible owing to the idea of self--similarity 
which our approach is based on and due to the convenient introduction of 
control functions through the algebraic transformation. The idea of 
self--similarity, complimented by the property of algebraic invariance, 
eliminates the umbiguity typical of divergent series in the standard 
perturbative approaches. In the sections that follow, it will be shown that
the accuracy of so constructed self-similar approximations is rather good.

Note that achieving good accuracy with a limited number of terms of an
asymptotic expansion should not be treated as surprising. Asymptotic series
are known to provide reasonable accuracy when up to some optimal number of
terms are taken [24], the subsequent terms only spoil the picture being, in
this sense, excessive. Whether there are such excessive terms or not is
decided, in our approach, by stability and crossover conditions. As soon as
these are satisfied, there are no excessive terms. And if adding more terms
does not allow us to satisfy these conditions, the added terms are to be
considered excessive. Fortunately, the real life and realistic physical
problems are so complicated that we practically never have excessive terms,
but vice versa, have to deal with very short expansions containing only a
few terms.

\section{Zero--dimensional model}

For illustrative purpose, we start with a simple model example.
Consider the partition function of a zero-dimensional anharmonic model, 
represented by the integral 
\begin{equation}
\label{21}
J(g)=\frac{1}{\sqrt{\pi}} \int_{-\infty}^\infty \exp \left (
- x^2 - gx^4 \right ) dx, 
\end{equation}
with the integrand possessing a single "vacuum" state, located at the
point $x=0$. The weak--coupling expansion of this integral in powers of
the coupling parameter $g$, around the vacuum state, leads to divergent
series, 
\begin{equation}
\label{22}
J(g) \simeq a + bg + cg^2 + \ldots \; , \qquad (g\rightarrow 0), 
\end{equation}
where
$$
a=1\; , \qquad b = - \frac{3}{4}\; , \qquad c = \frac{105}{32}\; . 
$$
The so-called strong-coupling expansion, in inverse powers of the
coupling constant, can be written down as well: 
\begin{equation}
\label{23}
J(g) \simeq Ag^{-1/4} + Bg^{-3/4} + Cg^{-5/4} + \ldots \; ,\qquad 
(g\rightarrow \infty ) \; , 
\end{equation}
with
$$
A= \frac{1.813}{\sqrt{\pi}}\; ,\qquad B = -\frac{0.612}{\sqrt{\pi}}\; ,
\qquad C = \frac{0.227}{\sqrt{\pi}} \; . 
$$
Following the approach of Section 2, one can derive the right crossover
approximation, 
\begin{equation}
\label{24}
J^*(g,\infty) = aA \left ( A^2 + a^2g^{1/2} \right )^{-1/2}, 
\end{equation}
with the right crossover index $s=1/2$ and crossover time 
$\tau=-A^3/(2a^2B)=1.55$. At $g=1$, the percentage error of formula (24), is
equal to $-7.38\%$, while the maximal error is reached at $g=0.35$ and 
equals $-7.96\%$.

The left crossover approximation is given as follows: 
\begin{equation}
\label{25}
J^*(g,0) = aA \left ( A^4 + a^4g \right )^{-1/4}, 
\end{equation}
with the left crossover index $s=1/4$, and crossover time 
$\tau=-a^5/(4A^4b)=0.304$. At $g=1$, the percentage error of Eq. (25) 
is $10.13\%$, while the maximal error at $g=2.5$, is equal to $10.53\%$. 
We conclude, that the crossover approximations (24) and (25) may be viewed,
correspondingly, as the lower and upper bounds for the integral (21). The
average, defined by Eq. (15),
$$
J^{*}(g) = \frac{J^*(g,0) + J^*(g,\infty)}{2}, 
$$
possesses the correct leading asymptotes and approximates the exact result
at $g=1$ with the percentage error of $1.37\%$. And the maximal error, at 
$g=3$, is $2.21\%$.

\section{Lattice gauge model}

The vacuum energy density $f_0$ of the (3+1)--dimensional SU(2) lattice gauge
model in its weak--coupling, asymptotically free regime, may be presented in
the form of an expansion in powers of the parameter $x=4/g^4$, where $g$
stands for the coupling [25]: 
\begin{equation}
\label{26}
f_0 \simeq Ax + B\sqrt{x}+ \ldots \; ,\qquad 
\left ( x\rightarrow \infty \right ) \; , 
\end{equation}
where
$$
A = -6,\qquad B = 7.1628\; . 
$$
In its strong--coupling limit, $f_0$ can be presented as follows [25]: 
\begin{equation}
\label{27}
f_0 \simeq ax^2 + bx^4 + \ldots \; ,\qquad 
\left ( x \rightarrow 0 \right )\; , 
\end{equation}
with
$$
a = - 1 \; ,\qquad b = 0.03525\; . 
$$
Because of the interfering roughening transition, the quality of the
high--order terms in the strong--coupling expansion is doubtful [25], so we
use only its leading terms. The left crossover approximation can be readily
written down as
\begin{equation}
\label{28}
f^{*}(x,0) = ax^2 \left [ 1 + \left ( \frac{a}{A} \right )^2x^2
\right ]^{-1/2}\; , 
\end{equation}
where we have used
$$
s = \frac{1}{2} \; , \qquad \tau = - \frac{a^3}{2bA^2}\; . 
$$
The numbers generated by formula (28), practically coincide in the region 
$x\in [0.1,1.1]$ with estimates obtained in [25] from the strong--coupling
approximants. The right crossover approximation can be obtained as well, but
its accuracy is worse than that of Eq. (28).

\section{One--dimensional Bose system}

The ground--state energy of the one--dimensional Bose system with the
$\delta$--functional repulsive interaction potential is known in a 
numerical form from the Lieb--Liniger exact solution [26]. It is 
desirable, nevertheless, to have a compact analytical expression for the 
ground--state energy $e(g)$ as a function of the $\delta$--function 
strength $g$, valid for arbitrary $g$. In the weak--coupling limit an 
exact analytical result is known: 
\begin{equation}
\label{29}
e(g) \simeq g\; , \qquad \left ( g\rightarrow 0\right ) \; , 
\end{equation}
obtained in Ref. [26], while in the strong--coupling limit another exact 
result, obtained by Girardeau [27], is available: 
\begin{equation}
\label{30}
e(g)\simeq \frac{\pi^2}{3} \; , \qquad 
\left( g\rightarrow \infty \right) . 
\end{equation}
The higher--order terms in these expansions were derived by approximate
methods, the next term in the weak--coupling limit being $\approx bg^{3/2}$,
and in the strong--coupling limit $\approx Bg^{-1}$. We shall not use the
approximate values for the coefficients, $b$ and $B$ (see e.g. [13]
and references therein), writing instead trial expansions and determining
the coefficients by matching the two asymptotic forms for the ground state
energy. Following the standard approach of Section 2, we obtain the right
crossover approximation: 
\begin{equation}
\label{31}
e^{*}(g,\infty) = \frac{\pi^2}{3} g \left( g+\frac{\pi^2}{3} 
\right)^{-1}, 
\end{equation}
with the right crossover index $s=1$ and $B=(\pi^2/3)^2$. Although, Eq. (31) 
can be further simplified, we leave it in present form in order to stress 
the origin of its different parts. Simple expression (31) works with 
surprising accuracy of about $1-2\%$, up to $g\sim 10$, till there are 
numerical data available for comparison [26]. The left crossover--type 
expression can be written as well, following the standard procedure, but 
its accuracy is inferior to that of Eq. (31).

\section{One--dimensional ferromagnet}

Low--dimensional magnetic systems give a plenty of examples of the crossover
phenomena, when only the asymptotic behavior with respect to different
parameters, such as spin, temperature etc, is known and the intermediate
region, in most of the cases, could be reached only numerically. The
crossover self--similar approximations offer simple analytical expressions
for the intermediate region. We put below, for simplicity, the exchange
integral $J=1$.

\subsection{Zero--field thermodynamics}

The free energy $F$ and magnetic susceptibility $\chi$ of the
one--dimensional Heisenberg ferromagnet of spin $S$, within the
framework of the spin-wave approximation, valid at temperatures 
$T\rightarrow 0$, has the form of an expansion in powers of $T$ [28]: 
\begin{equation}
\label{32}
F \simeq a(S) T^{3/2} + b(S) T^2 + \ldots \; ,\qquad (T\rightarrow 0)\; , 
\end{equation}
in which
$$
a(S) = -\frac{\zeta(3/2)}{(2\pi)^{1/2}} \left ( \frac{1}{2S} 
\right)^{1/2} , \qquad b(S) = \frac{1}{4S^2}\; , 
$$
and
\begin{equation}
\label{33}
\chi \simeq A(S)T^{-2} + B(S)T^{3/2}, \qquad  ( T \rightarrow 0)\; , 
\end{equation}
where
$$
A(S) = \frac{8}{3}S^4 \; , \qquad 
B(S) = -A(S)\frac{3\zeta(1/2)}{\sqrt{2\pi}S} \; . 
$$
As $T\rightarrow \infty$, a different asymptotic behavior happens [29]: 
\begin{equation}
\label{34}
F \simeq - T\ln(1+2S) \qquad ( T\rightarrow \infty )
\end{equation}
and
\begin{equation}
\label{35}
\chi \simeq \frac{4S(S+1)}{3T} \qquad (T\rightarrow \infty ). 
\end{equation}
Applying the standard approach of Section 2, we obtain for the free energy 
the following left crossover approximation, corresponding to the left 
crossover index $s=1$, 
\begin{equation}
\label{36}
F^*(T,S) = a(S) \frac{T^{3/2}}{1-[b(S)/a(S)]\tau T^{1/2}} \; ,
\qquad \tau = \frac{a^2(S)}{b(S)\ln(1+2S)} \; , 
\end{equation}
and the expression for specific heat $C^*=-Td^2 F^*(T,S)/dT^2$ as
\begin{equation}
\label{37}
C^*(T,S) = -\frac{1}{4} T^{1/2}a^3(S) 
\frac{-3a(S)+b(S)\tau T^{1/2}}{[-a(S)+b(S)\tau T^{1/2}]^3}\; . 
\end{equation}
The position, height and spin--dependence of the maximum occurring in the
expression for $C^{*}(T,S)$ are in qualitative agreement with numerical
results for finite chains [30].

The left crossover approximation for the renormalized susceptibility is
\begin{equation}
\label{38}
\chi^* = \frac{A(S)}{T^2} \left[ 1+ \frac{B(S)}{2A(S)} \tau T^{1/2}\right]^2,
\qquad \tau = \frac{A(S)}{B(S)} \left[ 
\frac{16S(1+S)}{3A(S)}\right]^{1/2}, 
\end{equation}
with the left crossover index $s=-2$.
The expressions (36) and (38) are very accurate for $S=1/2$, where they
practically coincide with the results of a numerical solution of the
thermodynamic Bethe--ansatz equations [28]. 

\subsection{Spin waves at finite temperatures}

Variational theory, as applied at low temperatures [31], gives the
temperature--dependent expression for the spin--wave energy 
$\omega_k$ for $S=1/2$ in the form 
\begin{equation}
\label{39}
\omega_k = 2 Z(T)\left| \sin (k)\right| \; , \qquad Z(T)\simeq
\frac{1}{2} \pi \left[ 1-\frac{2}{3} \left( \frac{T}{2} \right)^2 \right] 
\qquad (T\rightarrow 0), 
\end{equation}
being at $T=0$ completely in agreement with the exact results [28]. In order
to find the behavior of $Z(T)$ at arbitrary $T$, we continue it from the
region of $T\rightarrow 0$ self--similarly, along the most stable trajectory,
with the crossover index $s$, determined by the condition of the minimum
of the multiplier [20-22]
$$
m(T,s) = 1-\frac{1}{6} T^2 \frac{1+s}{s}\; , 
$$
from where 
$$
s(T) = \frac{1}{6} T^2 \left( 1 - \frac{T^2}{6} \right)^{-1},
\qquad T < \sqrt{6} \; , $$
$$
s \rightarrow \infty\; , \qquad T \geq \sqrt{6} \; . 
$$
This gives the left crossover approximation 
\begin{equation}
\label{40}
Z^*(T) = \frac{1}{2} \pi \left( \frac{s(T)}{s(T)+T^2/6}\right)^{s(T)},
\qquad T < \sqrt{6}\; , 
\end{equation}
\begin{equation}
\label{41}
Z^*(T) = \frac{1}{2} \pi \exp (-T^2/6) \; ,\qquad T\geq \sqrt{6}\; . 
\end{equation}
Formulae (40) and (41) suggest that the spin waves should survive at least 
up to $T\sim\sqrt{6}$, and become exponentially "soft" above this temperature.
Note that in this particular case the left self--similar approximation
plausibly reconstructed the function for arbitrary temperatures, even not
knowing beforehand the asymptotic behavior at $T\rightarrow\infty$.

\subsection{Field--dependent part of free energy}

It is believed that the magnetic field--dependent part of the free 
energy of the one--dimensional Heisenberg ferromagnet is independent on 
spin and scales as $\rho=h/T^2$ ($h$ denotes the magnetic field), with 
the scaling function independent on the value of spin [32]. For the 
classical ferromagnet, both low and high field behavior of the field 
dependent part of the free energy $\delta F(\rho)$ are known [32-35] in 
the simple form: 
\begin{equation}
\label{42}
T^{-2} \delta F(\rho) \sim a\rho^2 + b\rho^4 \; ,
\qquad a = -\frac{1}{3}\; , \qquad b = \frac{11}{135} \qquad (\rho \ll 1), 
\end{equation}
\begin{equation}
\label{43}
T^{-2}\delta F(\rho) \sim A\rho + B\rho^{1/2} , \qquad A = -1 \; ,
\qquad B = 1 \qquad (\rho \gg 1). 
\end{equation}
The left crossover approximation is controlled by the crossover index 
$s=1/2$ and crossover time $\tau=-a^3/(2A^2B)=0.227$,
yielding: 
\begin{equation}
\label{44}
T^{-2}\delta F^*(\rho,0) = a\rho^2 \left [ 1 + \left( 
\frac{a}{A} \right)^2\rho^2\right] ^{-1/2} ,
\end{equation}
while the right crossover approximation is given by 
\begin{equation}
\label{45}
T^{-2}\delta F^*(\rho,\infty) = A\rho^2 \left[ 
\rho^{1/2} + \left( \frac{A}{a} \right)^{1/2} \right]^{-2} , 
\end{equation}
with 
$$
s=2 \; , \qquad \tau=-2\frac AB\left(\frac{A}{a}\right)^{1/2}=3.464 \; .
$$
Both expressions (44) and (45) are in good agreement with the known results
[32-35].

\section{Flexible polymer coil}

The calculation of the so--called expansion function $\alpha^2(z)$ of a 
flexible polymer coil is of long standing interest in polymer science
[36-38]. This quantity defines the ratio of the mean square end--to--end 
distance $< R^2>$ of the chain to its unperturbed value $< R^2>_0\equiv 
Nl^2$, where $N$ is the number of segments with the length $l$ each, so 
that $Nl$ is the contour length of the chain,
\begin{equation}
\label{46}
\alpha^2(z) \equiv \frac{<R^2>}{<R^2>_0} \; ,
\end{equation}
as a function of a dimensionless interaction parameter $z$. The latter
is
\begin{equation}
\label{47}
z \equiv \frac{BN}{\pi l^2} \qquad ( D=2)
\end{equation}
for the two--dimensional case and
\begin{equation}
\label{48}
z =\left (\frac{3}{2\pi}\right )^{3/2} \frac{B\sqrt{N}}{l^3}
\qquad ( D=3)
\end{equation}
for the three--dimensional coil, where $B$ is the effective binary 
cluster integral for a pair of segments.

When the excluded volume interaction is very weak, a perturbation theory 
leads [39] to an asymptotic series
\begin{equation}
\label{49}
\alpha^2(z) \simeq 1 + \sum_{n=1} a_n z^n \qquad (z\rightarrow 0) \; ,
\end{equation}
in which the coefficients for the two--dimensional case are
$$
a_1 =\frac{1}{2} \; , \qquad a_2 = - 0.121545 \; , \qquad
a_3 = 0.026631 \; , \qquad a_4 =-0.132236 \qquad ( D=2)\; ,
$$ 
and for the three--dimensional coil they take the values
$$
a_1 =\frac{4}{3} \; , \qquad a_2 = - 2.075385 \; , \qquad
a_3 = 6.296880 \; , \qquad a_4 =-25.057251 \; ,
$$ 
$$
a_5=116.134785 \; , \qquad a_6=-594.71663 \qquad (D=3) \; .
$$
The asymptotic result for the strong coupling limit [40] is
\begin{equation}
\label{50}
\alpha^2(z) \simeq A_1 z^\beta + A_2 z^\gamma \qquad
(z\rightarrow\infty) \; .
\end{equation}

Using our method of self--similar interpolation, we obtain from (49) and (50)
\begin{equation}
\label{51}
\alpha_*^2(z) = \left ( 1 + A_1^{1/\beta} z\right )^\beta
\end{equation}
in the first approximation. The second approximation gives
\begin{equation}
\label{52}
\alpha_*^2(z) =\left [ \left ( 1 + C_1z\right )^{2-\beta+\gamma} +
C_2 z^2\right ]^{\beta/2} ,
\end{equation}
where
$$
C_1 = \left ( \frac{2A_2}{\beta A_1^{1-2/\beta}}
\right )^{1/(2-\beta+\gamma)} , \qquad C_2 = A_1^{2/\beta} .
$$
These formulae can serve for both the two-- as well as for 
three--dimensional coils. We shall concentrate on the latter case for which
accurate numerical data for $\alpha^2(z)$ are available [40] in the whole 
range of $z\in [0,\infty)$. Then in the strong coupling limit (50), one has
\begin{equation}
\label{53}
A_1=1.5310 \; , \qquad A_2=0.1843\; , \qquad \beta=0.3544\; ,
\qquad \gamma=-0.5756 \qquad (D=3) \; .
\end{equation}
The coefficients in (52) become
$$
C_1=6.5866 \; , \qquad C_2=11.0631\; , \qquad
2 - \beta + \gamma = 1.07 \; .
$$
In this way, from (52) we obtain
\begin{equation}
\label{54}
\alpha_*^2(z) =\left [ \left ( 1 + 6.5866 z\right )^{1.07} +
11.0631 z^2 \right ]^{0.1772} .
\end{equation}
The self--similar approximation (54) is accurate, within $0.4\%$ of error,
in the full range $z\geq 0$, as compared to numerical calculations [40]. This
formula (54) practically coincides with the phenomenological extrapolation
expression
\begin{equation}
\label{55}
\alpha_{MN}^2(z) = \left ( 1 + 7.524 z + 11.06 z^2\right )^{0.1772} ,
\end{equation}
obtained by Muthukumar and Nickel [40] by means of a fit to numerical data.

\vspace{3mm}

In conclusion, we have developed the {\it method of self--similar 
interpolation} for deriving explicit interpolation formulae for difficult
crossover problems of statistical mechanics. This method, as is 
illustrated by several examples, is general, simple, and accurate.

\end{sloppypar}

\vspace{3mm}

{\bf Acknowledgement}

\vspace{2mm}

We appreciate useful discussions with E.P. Yukalova.

\end{document}